\documentclass[9pt,conference]{IEEEtran}
\IEEEoverridecommandlockouts

\usepackage{cite}
\usepackage{amsmath,amssymb,amsfonts}
\usepackage{booktabs}
\usepackage{mathrsfs}
\usepackage{algorithmic}
\usepackage{graphicx}
\usepackage{textcomp}
\usepackage{xcolor}
\usepackage[colorlinks, citecolor=blue]{hyperref}

\usepackage{CJKutf8}

\def\BibTeX{{\rm B\kern-.05em{\sc i\kern-.025em b}\kern-.08em
    T\kern-.1667em\lower.7ex\hbox{E}\kern-.125emX}}
\begin{document}

\title{CP-UNet: Contour-based Probabilistic Model for Medical Ultrasound Images Segmentation}

\author{
    \IEEEauthorblockN{Ruiguo Yu$^{1,2,3,4}$, Yiyang Zhang$^{2,3,4}$, Yuan Tian$^{1,2,3}$, Zhiqiang Liu$^{1,2,3}$, Xuewei Li$^{1,2,3,4}$, Jie Gao$^{1,2,3,\ast}$\thanks{$^{\ast}$ Corresponding author.}}
    \IEEEauthorblockA{$^1$ College of Intelligence and Computing, Tianjin University, Tianjin, 300350, China.}
    \IEEEauthorblockA{$^2$ Tianjin Key Laboratory of Cognitive Computing and Application, Tianjin, 300350, China.}
    \IEEEauthorblockA{$^3$ Tianjin Key Laboratory of Advanced Networking, Tianjin, 300350, China.}
    \IEEEauthorblockA{$^4$ School of Future Technology, Tianjin University, Tianjin, 300350, China.}
    \IEEEauthorblockA{ \{rgyu, zyy\_0203, tiany, tjubeisong, lixuewei, gaojie\}@tju.edu.cn}
}

\maketitle
\begin{abstract}

Deep learning-based segmentation methods are widely utilized for detecting lesions in ultrasound images. Throughout the imaging procedure, the attenuation and scattering of ultrasound waves cause contour blurring and the formation of artifacts, limiting the clarity of the acquired ultrasound images. To overcome this challenge, we propose a contour-based probabilistic segmentation model CP-UNet, which guides the segmentation network to enhance its focus on contour during decoding. We design a novel down-sampling module to enable the contour probability distribution modeling and encoding stages to acquire global-local features. Furthermore, the Gaussian Mixture Model utilizes optimized features to model the contour distribution, capturing the uncertainty of lesion boundaries. Extensive experiments with several state-of-the-art deep learning segmentation methods on three ultrasound image datasets show that our method performs better on breast and thyroid lesions segmentation.  

\end{abstract}


\section{Introduction}
\begin{figure}
    \centering
    \includegraphics[scale = 1, trim=0cm 0.1cm 0cm 0cm, clip,width=1\linewidth]{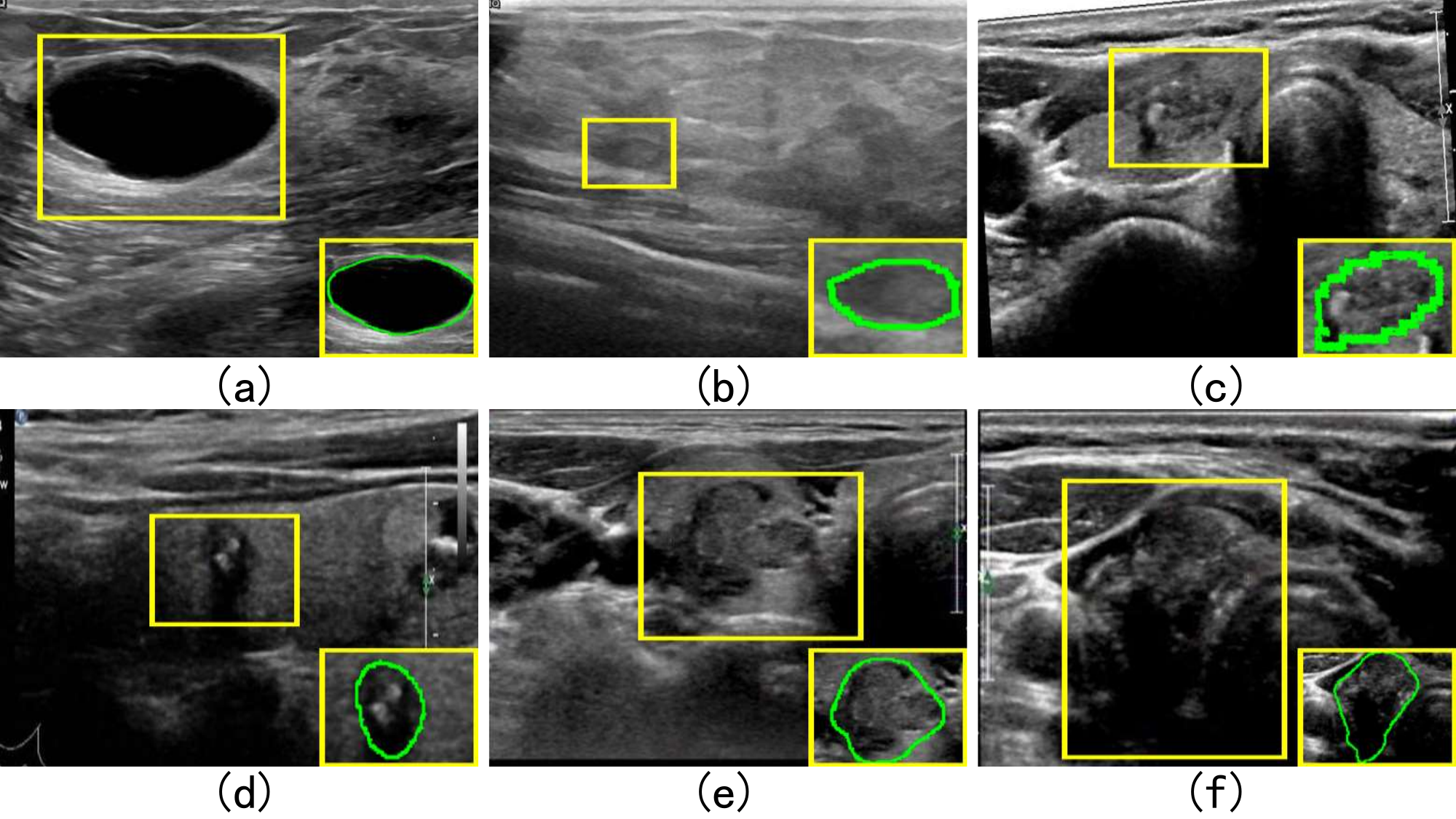}
    \caption{Enlarged images of nodules in six ultrasound images: (a) Clear contours; (b)-(d) Blurred contour edges; (e)-(f) Irregular shapes. The green contour line is physician-labeled}
    \label{clarification}
\end{figure}

Ultrasound imaging is widely used for various disease diagnoses due to its low cost, simple operation, and non-invasiveness. With the development of computer-aided diagnosis (CAD) technology, deep learning-based segmentation methods have been applied to lesion segmentation in medical ultrasound images. However, inhomogeneous lesion region distribution, speckle noise, and imaging artifacts in ultrasound images increase the difficulty of the segmentation task. Structural boundary line ambiguity, such as the unclear transition between lesions and normal tissues as shown in Fig.~\ref{clarification}(b)(c)(d), and contour heterogeneity, including irregular variations in lesion shapes and textures illustrated in Fig.~\ref{clarification}(e)(f), both contribute to contour blurring. These factors collectively affect the accuracy of segmentation.

Existing approaches primarily focus on optimizing convolutional neural networks to enhance the accuracy of medical image segmentation.UNet\cite{ronneberger2015u} and its variants\cite{zhou2018unet++}\cite{zhang2018road}\cite{8697107}effectively integrate features at different scales through an encoder-decoder architecture, which enables the network to extract fine-grained local features from the image. However, A limited receptive field limits the perception of detailed features by convolutional operations and can only process local information within the size of the convolutional kernel. To alleviate the limitations of convolutional operations, ReAgU-Net \cite{ding2019automatic} and UNeXt \cite{valanarasu2022unext} propose different strategies to expand the receptive field. While enhancing the contextual connection of the deep encoding stage, the shallow encoding stage still does not introduce the global connection of detailed features, which affects the accurate recognition of blurred contours. Other methods based on the attention mechanism, such as ConTrans\cite{10.1007/978-3-031-16443-9_29},  ACC-Unet \cite{ibtehaz2023acc}, Segnetr \cite{cheng2023segnetr},  utilize the improved self-attention mechanism to establish connections between detail features and enhance the global local connections in the full-stage encoding. However, they need to develop interconnections between all features, which is unnecessary for the sparsity of contours in ultrasound images. We utilize the fusion strategy in the attention mechanism to establish global links that fuse sparse features while preserving detailed features.

\begin{figure*}
    \centering
    \includegraphics[scale = 0.55, trim=0.2cm 3.43cm 0.6cm 1.3cm, clip,width=1\linewidth]{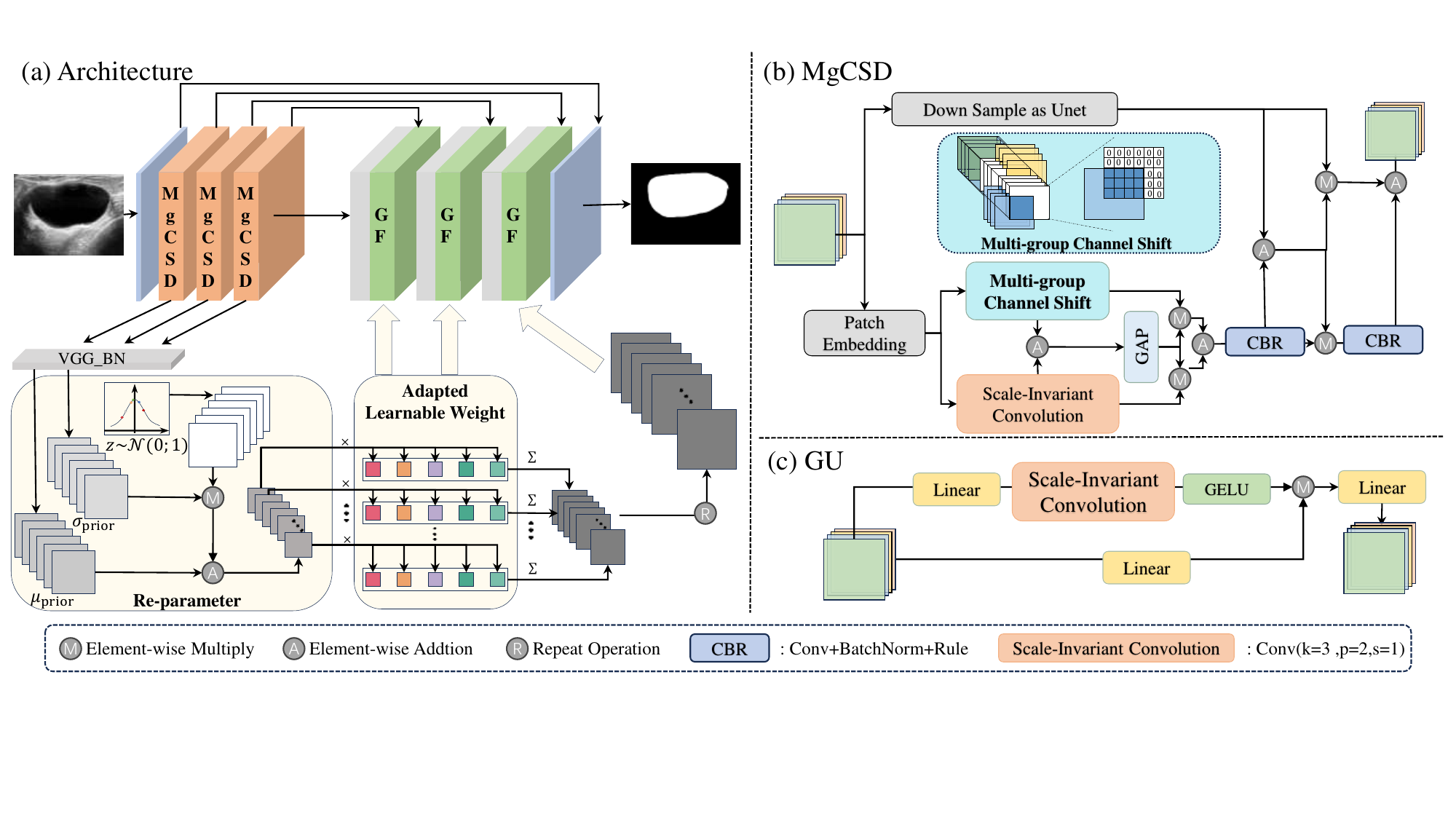}
    \caption{Framework of CP-UNet. CMP, Contour Probabilistic Modeling. MgCSD, Multi-group channel shifted downsampling. GF, Gating-based feature filtering module}
    \label{arch}
\end{figure*}

There are many methods to create constraints specifically for the contour features to guide the network in enhancing its focus on the contours to address the domain of ambiguous contours in ultrasound images. For example, GGNet\cite{xue2021global} and Wavelet U-Net++\cite{agnes2024wavelet} both provide constraints on the extent of the lesion region for the encoded intermediate features. AGMT\cite{zhao2024learning} investigates a transformer-based breast ultrasound loss function that enhances the perception of the lesion's shape. However, the loss function cannot fully express the contour features to learn the specific features of the contour. When the local edge line of the contour is blurred, it needs to be complemented by the global edge line characteristics, such as pixels around the edge line, contour shape, and range characteristics. We use the Gaussian Mixture Model to simulate the distributional features of the synthetic contours to break the limitation of a single loss constraint and establish generalized features to guide the contour representation in the segmentation network.

To obtain generalized contour features to guide the segmentation process, we propose CP-UNet, in which the Multi-group Channel Shift Downsampling (MgCSD) utilizes a fusion strategy in the attention mechanism to give global connections to preserve detailed features and provide more comprehensive features. Then, the Contour Probabilistic Modeling Module (CPM) in the network breaks the limitation of a single loss constraint on the contour to build generalized contour features. Finally, the gating mechanism-based feature filtering module (GF) provides effective fusion strategies for contour features.

The contributions of this paper are summarized as follows:

\begin{itemize} 
\item[$\bullet$] We propose a contour probabilistic modeling-guided ultrasound image segmentation network, which fits the contour features with probability distributions and guides the enhancement of the contour representation of the features at different stages to improve the segmentation results. 
\item[$\bullet$] MgCSD is proposed instead of traditional convolutional downsampling to create global-local links for full-stage coding features. GF can fuse contour features into the decoding stage, bridging the semantic gap between MgCSD and CPM and downsampling results.
\item[$\bullet$] Extensive experiments on BUSI \cite{ALDHABYANI2020104863}, DDTI \cite{10.1117/12.2073532}, and one private thyroid ultrasound images dataset demonstrate that our approach consistently improves the segmentation accuracy of breast nodules and thyroid nodules, outperforming the strong baseline and state-of-the-art medical image segmentation methods. 
\end{itemize}

\section{Method}
The CP-UNet architecture proposed is shown in Fig.~\ref{arch}(a). Firstly, the image is downsampled using multiple multi-group channel shift downsampling (MgCSD) modules. Secondly, the results of MgCSD are fed into the contour probabilistic modeling (CPM) module to fit the contour features of the original image using the mixed Gaussian distribution to fit the contour features. Finally, the gating-based feature filtering module (GF) fuses the up-sampling result, the down-sampling consequence, and the CPM sampling result of the same dimension at the corresponding stage. 
Finally, the nodule's segmentation result is obtained. The following subsections will introduce the MgCSD, CPM, and GF modules.

\subsection{Multi-group channel shifted downsampling}
\label{section: A}
Shallow convolutional layers can effectively capture localized edge line features but often lack a grasp of contour structure. Deeper convolutional layers, expanding the receptive field and capturing the contextual relationships of contour features, may lose detailed representations like edge lines. We propose a downsampling method based on spatial shift weighting of multiple groups of channels.

The shift operation is shown in Fig.~\ref{arch}(b), where the features are divided into groups according to the channel dimensions. The first group remains fixed, and the rest are shifted across the width and the height according to cyclic upward and cyclic rightward shifts, leaving the relative positions of the elements in the regions. The non-feature areas are filled with zeros after the shifts. The shifted features are denoted by $F_{shift}$.

The shift operation links edge regions with central features but results in the loss of detailed information from the areas displaced. So we use scale-invariant convolution to add detailed information, denoted as $F_{supply}$.

Inspired by the channel self-attention module\cite{chen2022aau}, this module enhances the performance of $F_{shift}$ and $F_{supply}$ in terms of the channel dimensions by utilizing the result of global average pooling, which is fused to obtain the globally enhanced weights $F_L$, which is represented as follows:
\begin{equation}
    F_L = GAP(F_{shift} \oplus F_{supply}) \otimes  (F_{shift} \oplus F_{supply})
\end{equation}
where $ \otimes $ and $ \oplus $ are element-wise operation

Complete local feature preservation is provided by convolutional downsampling, denoted as $F_R$.The result of adding $F_L$ and $F_R$ will be mapped to obtain an interrelationship matrix $ O_{mixer}$. For the $F_R$, $ O_{mixer} $ preserves the global features with the spatial features of the right branch. The weighting is the same as that of the spatial self-attention module in \cite{chen2022aau}, and the fusion process is represented as follows:
\begin{equation}
    F = O_{mixer} \otimes  F_{L} \oplus O_{mixer} \otimes  F_{R}
\end{equation}
where $ O_{mixer},F_{L},F_{R} \in {\mathbb{R}}^{C \times \dfrac{h}{2} \times \dfrac{w}{2}}$, $ \otimes $ and $ \oplus $ are element-wise operation

MgCSD replaces and enhances the traditional downsampling module and can be used as a direct replacement in the encoding phase. In particular, note that since the global weighted branch targets feature rather than the original image, the simple convolution of the first layer is retained when used to map the original image into the feature space.

\subsection{Contour Probabilistic Modeling}

Due to the individual variability in the presentation of a lesion's contour, when the contour's local edge lines are unclear, downsampling in relation to the global can be supplemented with contour information from other regions. When the shape of the contour is irregular, supplementing local edge lines requires consideration of global contour variations. We use probabilistic modeling to fit the generalized contour features, representing the shape of the contour, the edge line, the degree of blurring, and other information. In contrast, the Gaussian mixture distribution can be fitted to any distribution. So, the probability density function of the hybrid Gaussian distribution can be expressed as several Gaussian distributions of the linear weighting of the weights and the sum of the values of one. 

Taking advantage of the Gaussian mixture distribution, we assume that the K sets of Gaussian distribution features of the contour are learned from the original image X, denoted as $ \mu_{A}( \textit{X};\theta), \sigma_{A}(\textit{X};\theta) $. 

Specifically, the output of MgCSD as the input of the CPM, denoted as ${\left\{ c_{i} \right\} }_{i=1}^{L}$, L represents the number of encoder layers. It is then mapped by a feature extractor to the K sets of Gaussian distributions, denoted as $ \mu_{A}( \textit{X};\theta), \sigma_{A}(\textit{X};\theta) = {\left\{ \mu_{i} ( \textit{X};\theta), \sigma_{i}( \textit{X};\theta) \right\} }_{i=0}^K $. 

To fuse the obtained contour features into the decoding stage, we use the reparameterization trick \cite{kingma2013auto} to sample the obtained K sets of Gaussian distributions. As shown in Fig.~\ref{arch}(a), the sampling result is weighted and summed with the adaptive learnable weights $\Omega$ to achieve the purpose of sampling from the mixed Gaussian distribution. The whole process is called adaptive homoscedastic resampling and is expressed as follows: \begin{equation} G = \Omega ( z \cdot \sigma_{A} + \mu_{A}) \end{equation} where $\Omega \in {\mathcal{R}}^{T \times K} $,$T$ corresponds to the number of eigen channels in the corresponding stage. z is sampled from the standard normal distribution $\mathcal{N}(0, 1) $. 

Since the image contour is not only a closed geometric image, the pixel features around the contour line are also crucial for constructing the contour distribution, so the result from the mask-processed image Y mapped by the same feature extractor as CPM can be used,  which is expressed as $ \mu_{B}( \textit{Y};\theta'), \ sigma_{B}( \textit{Y};\theta') = {\left\{ \mu_{i}( \textit{Y};\theta'), \sigma_{i}( \textit{Y};\theta') \right\} }_{i=0}^K $. 

To constrain the contour distribution to fit the actual distribution of the contour and focus on the critical area \cite{Lei_2024_CVPR}, the KL dispersion will align $ \mu_{A}, \sigma_{A}$ toward $ \mu_{B}, \sigma_{B}$, which is denoted as:
\begin{equation}
    KL_{AB} = \log{\dfrac{\sigma_{B}}{\sigma_{A}}}+\dfrac{\sigma_{A}^{2}+(\mu_{A}-\mu_{B})^2}{2\sigma_{B}^{2}}-\dfrac{1}{2}
\label{KLab}
\end{equation}
where $ \mu_{A}, \sigma_{A} $ and $ \mu_{B}, \sigma_{B}$, are both composed of K sets of Gaussian distributions, each of which is a pair of combinations of mean-variance values.

\subsection{Gating-based feature filtering module}

As shown in Fig.~\ref{arch}(c), to fuse the multiple data inputs sampled on each stage, a gating mechanism filters the input features $F \in {\mathbb{R}}^{n \times h \times w },$ such that important features are retained and irrelevant or noisy features are suppressed. F is denoted as follows:
\begin{equation}
F = F_{up} \oplus F_{skip} \oplus F_{contour}
\label{Ffuse}
\end{equation}
where $ F_{up} \in {\mathbb{R}}^{c \times h \times w} $ is the result of the sampling on the previous layer of decoder result, $ F_{skip} $ and $ F_{contour} $ are the coded features of the corresponding stage and the contour features derived from the adaptive resampling of the same source, respectively, with the same size as $ F_{up} $.

The GF module establishes channel-based gating branches to create dynamically selected intermediate representations for high-frequency information $ F $. The gating branch is established using the linear mapping Linear, the scale-invariant convolution, and the activation function GELU, and the gating signals $ G $ obtained all contribute to the hidden units from the feature information of different channels at the exact spatial location. Finally, the gating signal weights are weighted to each embedded channel feature using the scale operation, which can be expressed as:
\begin{equation}
    F_{gate} = G^T \otimes Linear(F)
\end{equation}
where $ G\in {\mathbb{R}}^{c \times n}, n = h \times w$, $h$ and $w$ are consistent with the size of the feature $F$, $n$ is consistent with the embedding dimension of the $ Linear(F) $ operation, and $ G^T $ is the result of the transposition of $G$.$F_{gate}$will serve as the input for the next stage of upsampling.

\section{Experiments}
\subsection{Datasets and Experimental Settings } 
This paper validates the model in this paper on two types of ultrasound datasets, breast and thyroid. The breast ultrasound dataset BUSI\cite{ALDHABYANI2020104863} removed 133 images of standard cases without breast lesions, using 665 images containing nodules. The thyroid ultrasound dataset DDTI\cite{10.1117/12.2073532} removes labeled corrupted images and expands images with multiple nodules, using 872 images. The experiments are supplemented with the private thyroid ultrasound dataset TUI, which contains 15,233 images, to validate the results thoroughly.

The entire network is optimized using the SGD algorithm with a momentum of 0.9, weight decay of 0.01, batch size of 64, and epoch 200. This paper's initial learning rate is set to 1e-3 and is scaled down with the training process using the cosine annealing algorithm. The network is trained on a single RTX-3090 GPU. We use VGG16BN as a feature extractor in CPM. We use a combination of binary cross entropy (BCE), dice loss, and KL dispersion loss to train CP-UNet, expressed as:
\begin{equation}
    \mathscr{L} = BCE(\hat{y},y) + Dice(\hat{y},y) + KL_{AB}
\end{equation}
where y is the predicted outcome, $\hat{y}$ is the target outcome, and $ KL_{AB} $ is the result of \eqref{KLab}.

\subsection{Performance Comparison} 

In this paper, we compare the performance of CP-UNet with widely used medical image segmentation methods, which include convolutional baselines such as UNet\cite{ronneberger2015u}, UNet++\cite{zhou2018unet++}, ResUNet\cite{zhang2018road}; SETR\cite{Zheng_2021_CVPR}, TransUNet\cite{chen2021transunet}, an improved segmentation network based on transformer, UNETR\cite{hatamizadeh2022unetr}, MedT\cite{valanarasu2021medical}; and Attention-based medical image methods\cite{ibtehaz2023acc, cheng2023segnetr,xue2021global,chen2022aau,valanarasu2021medical}. To make a fair comparison, this paper downloads the code of their public implementations and retrains them on all datasets.

\begin{figure}
    \centering
    \includegraphics[scale = 1, trim=1.4cm 2.8cm 1cm 0.6cm, clip,width=1\linewidth]{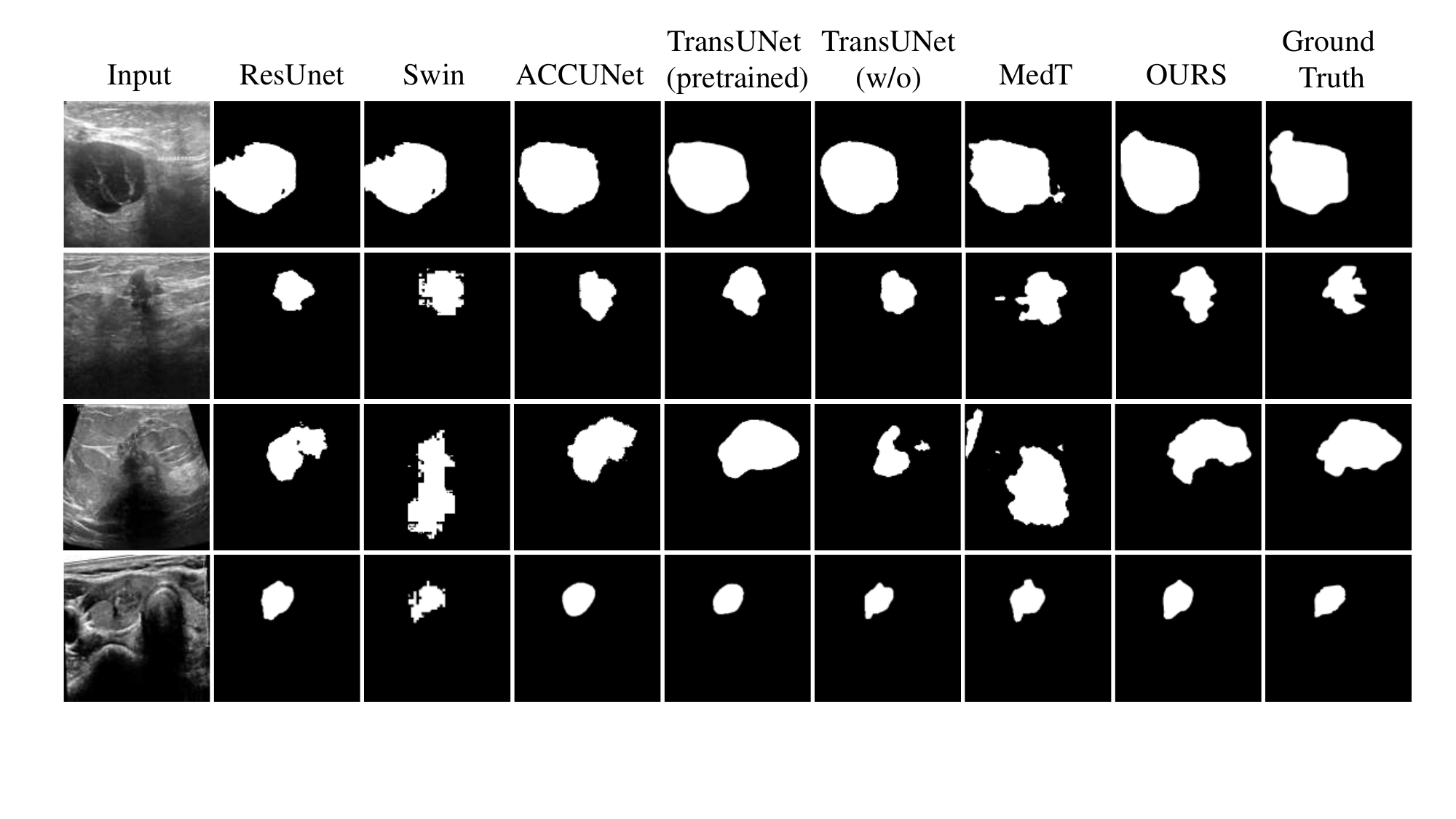}
    \caption{Visualization of the results on two ultrasound images. The first three rows show the segmentation results of the breast ultrasound image, and the last row of the thyroid ultrasound}
    \label{result}
\end{figure}

Fig.~\ref{result} shows the results of the segmentation visualization on the three datasets, and it can be seen that our method exhibits a relatively average level of regional delineation of the nodes, but for the trend direction of the contours, the performance is excellent and close to that of the truth map.

\begin{table}[t]
\caption{Performance Comparison with Other Method} 
\centering
\label{1-table} 
\resizebox{\linewidth}{!}{
\begin{tabular}{ccccccc} 
\toprule[1.5pt]
& \multicolumn{2}{c}{BUSI}  & \multicolumn{2}{c}{DDTI}  & \multicolumn{2}{c}{TUI}   \\
\cmidrule(r){2-3} \cmidrule(r){4-5} \cmidrule(r){6-7} & IoU(\%) & Dice(\%) & IoU(\%) & Dice(\%) & IoU(\%) & Dice(\%)  \\
\hline
UNet\cite{ronneberger2015u} & 0.5070 & 0.6729 & 0.6692 & 0.7999  & 0.7965 & 0.8866\\
UNet++\cite{zhou2018unet++} & 0.5935 & 0.7444 & 0.7299 & 0.8429  & 0.8034 & 0.8901\\
ResUNet\cite{zhang2018road} & 0.5517 & 0.7069 & 0.7121 & 0.8294   & 0.7631 & 0.8631\\
\midrule
SETR\cite{Zheng_2021_CVPR} & 0.2293 & 0.3720 & 0.2864 & 0.4434 & 0.2629 & 0.4141\\
MedT\cite{valanarasu2021medical} & 0.5155 & 0.6790 & 0.5873 & 0.7366  & 0.6382 & 0.7752  \\
Swin-Unet\cite{cao2022swin} & 0.4850 & 0.6503 & 0.4982 & 0.6602 & 0.3978 & 0.5636   \\
TransUNet(pretrained)\cite{chen2021transunet} & 0.6454 & 0.7812 & 0.8147 & 0.8963 & 0.8481 & 0.9166\\
TransUNet(w/o pretrained)\cite{chen2021transunet} & 0.5966 & 0.7448 & 0.7612 & 0.8625 & 0.8066 & 0.8909\\
\midrule
UNeXt\cite{valanarasu2022unext} & 0.5919 & 0.7428 & 0.7746 & 0.8722 & 0.8358 & 0.9104\\
Acc-Unet\cite{ibtehaz2023acc} & 0.5989 & 0.7463 &0.7483 & 0.8541   & 0.8255 & 0.8994\\
GGNet\cite{xue2021global} & 0.5410 & 0.7020  & 0.5594 & 0.7154 & 0.6744 & 0.8039 \\
AAU-net\cite{chen2022aau} & 0.6286 & 0.7714 & 0.7731 & 0.8704 & 0.7802 & 0.8754\\
ours & \textbf{0.6445} &\textbf{0.7827} & \textbf{0.8136} & \textbf{0.8972}  &  \textbf{0.8496} & \textbf{0.9183}\\
\bottomrule[1.5pt]
\end{tabular} 
}
\end{table}

\begin{figure}
    \centering
    \includegraphics[scale = 1, trim=0.4cm 1.8cm 1cm 2cm, clip,width=1\linewidth]{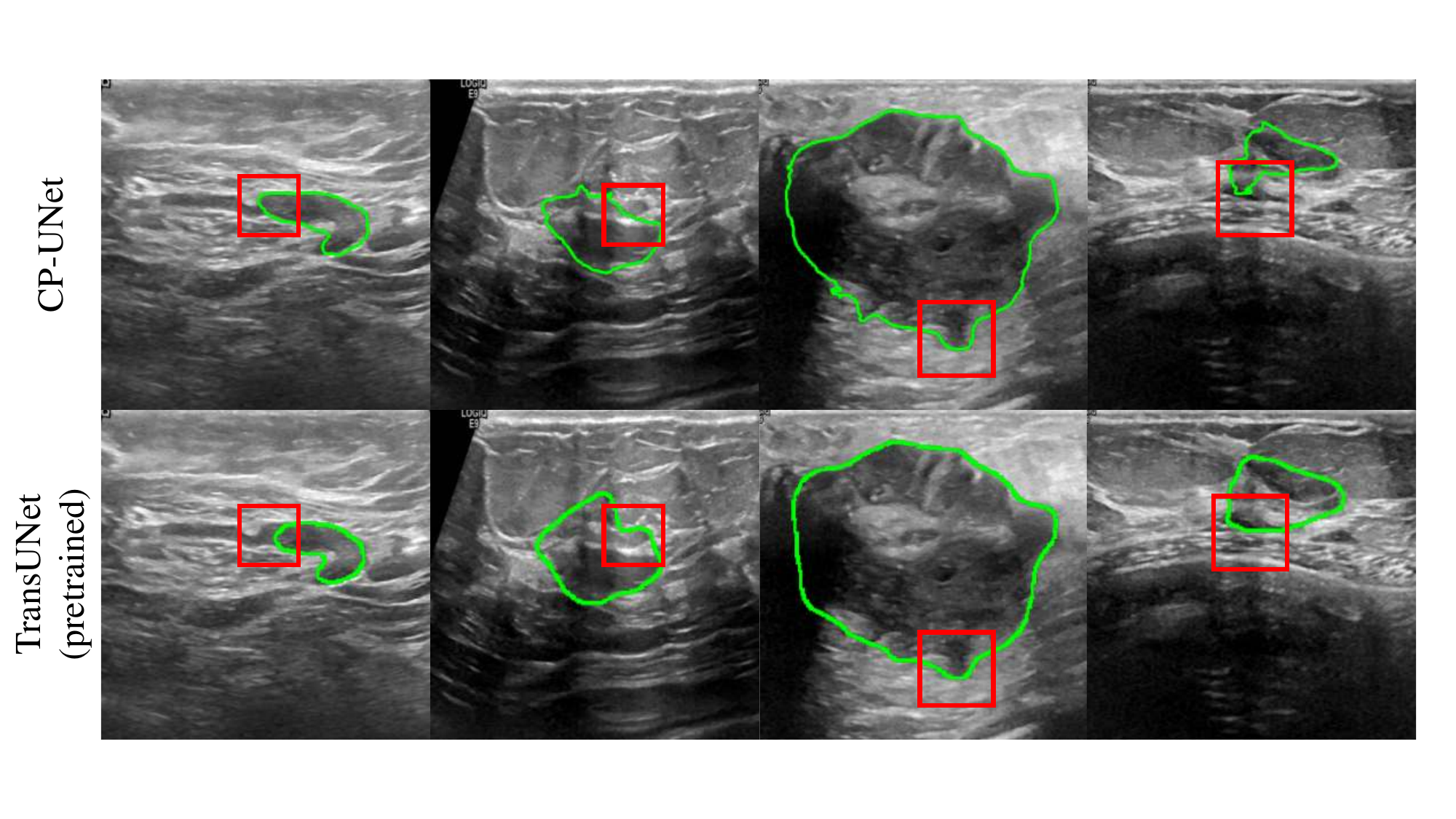}
    \caption{Visualization of contour segmentation performance by TransUNet and CP-UNet. The red boxes are poorly outlined areas, and the green line is the output of the model.}
    \label{visual}
\end{figure}

Table \ref{1-table} reports the experimental results of this paper's method and all competing methods on two ultrasound datasets, thyroid (DDTI, TUI) and breast (BUSI). On both the BUSI and DDTI datasets, the convolution-based baseline performs poorly compared to the model in this paper, which delivers about ten percent improvement and is comparable to the results of the pre-trained TransUNet on the ImageNet-21K dataset. Fig.~\ref{visual} shows that when the segmentation metrics are similar, CP-UNet is better at capturing contour zigzag variations than the pre-trained TransUNet, proving the effectiveness of our contour-specific modeling.

The experimental results in the last two columns of Table \ref{1-table} show that with sufficient data support, the attention-based method obtains a more stable enhancement and performs better than the convolutional baseline. At the same time, the model in this paper can achieve better results than the attention-based baseline, which indicates that the network proposed in this paper can build better support for the features in addition to the sufficient amount of data. Compared to the pure convolutional baseline, the method in this paper has a more significant improvement in BUSI than TUI and a considerable improvement over the attention-based baseline, indicating that the network proposed in this paper can extract more practical information for segmentation enhancement on a smaller dataset.

Fig.~\ref{visual} presents the segmentation results of ultrasound images obtained using different comparison methods. The first three rows display the segmentation results for breast nodules, while the last row shows the results for thyroid nodules. Except for the first row, the contours in the other images are unclear. Although the segmentation results of the pre-trained TransUNet appear similar to the ground truth, it fails to capture the intricate variations of the contours, rendering them smoother compared to our model. As shown in the figure, our model demonstrates superior performance in contour delineation.

\subsection{Ablation Studies} 
To validate the effectiveness of each module, we performed an ablation study on the BUSI dataset. Table \ref{melt-table1} shows the comparative results of the different components of the methodology in this paper. The baseline is constructed after culling all the network's MgCSD, GF, and CPM modules. Note that when culling the MgCSD, the network is missing encoders; at this point, we use the encoders in the unit to replace the MgCSD modules in each stage.

Adding all three modules alone improves the experimental effect relative to the baseline. MgCSD improves the model effect by about $10\%$, suggesting that the encoding strategy of mutual global-local feature enhancement is effective. The impact of the CPM module alone has a limited effect on the baseline enhancement in line with our expectations, as the CPM module needs the MgCSD to provide it with better quality features, as well as the GF module for its filtering, as evidenced by the experimental comparison of MgCSD+CPM and GF+CPM versus each module alone.GF+CPM is slightly less effective than GF alone. We do not rule out that it is due to the lack of information in CPM caused by the absence of MgCSD because GF+MgCSD+CPM is higher than GF+MgCSD. Together, the three modules can achieve the best results, suggesting that the GF feature fusion strategy is necessary for decoding the three data sources.

\begin{table}
\label{melt-table1} 
\caption{Performance of Ablation Study.}
\centering
\begin{tabular}{ccccc}
\toprule
MgCSD & GF & CPM  & IoU & Dice \\
\midrule
 - & - & -  & 0.5439 & 0.7038\\
\checkmark & - & - &  0.6410 & 0.7785\\
 - & \checkmark & -  & 0.5752 & 0.7299\\
 - & - & \checkmark  & 0.5550 & 0.7067\\
\checkmark & \checkmark & - &  0.6323 & 0.7739\\
 - &\checkmark & \checkmark  & 0.5713 & 0.7208\\
\checkmark & - &\checkmark  & 0.6420 & 0.7795\\
\checkmark & \checkmark & \checkmark  & \textbf{0.6445} & \textbf{0.7827}\\
\bottomrule
\end{tabular}
\end{table}

\section{Conclusion}
For focus blurred lesion contour, we propose a contour-based probabilistic modeling medical ultrasound image segmentation network (CP-UNet) for lesion segmentation in ultrasound images. Adaptive homologous resampling and hybrid Gaussian distribution modeling of global semantic features in the CPM module optimize the predicted contours to help improve segmentation performance. In this paper, the network is evaluated on various ultrasound images on public and private datasets and compared with state-of-the-art methods. In future work, we will explore the potential of contour-based attention mechanisms to improve the diagnostic accuracy for nodules.
\bibliographystyle{IEEEtran}
\bibliography{IEEEexample}
\end{document}